\documentclass[twocolumn]{aastex63}

\usepackage{amsmath}
\usepackage{booktabs}
\usepackage{threeparttable}
\usepackage{tabularx}

\usepackage{lineno}
\submitjournal{ApJL}

\shorttitle{Star Formation Law From BAO Intensity Mapping}
\shortauthors{Sun}
\graphicspath{{./}{figures/}}
\begin{document}

\defcitealias{BL_2011}{BL11}

\title{Cosmological Constraints on the Global Star Formation Law of Galaxies: Insights From Baryon Acoustic Oscillation Intensity Mapping}

\email{gsun@astro.caltech.edu}

\author{Guochao Sun}
\affiliation{California Institute of Technology, 1200 E. California Blvd., Pasadena, CA 91125, USA}

\begin{abstract}
Originally proposed as a cosmological probe of the large-scale structure, line intensity mapping (LIM) also offers a unique window into the astrophysics of galaxy evolution. Adding to the astrophysical explorations of LIM technique that have traditionally focused on small, non-linear scales, we present a novel method to study the global star formation law using forthcoming data from large-scale baryonic acoustic oscillation (BAO) intensity mapping. Using the amplitude of the percent-level but scale-dependent bias induced by baryon fraction fluctuations on BAO scales, we show that combining auto- and cross-correlation power spectra of two (or more) LIM signals allows to probe the star formation law power index $\mathcal{N}$. We examine the prospect for mapping H$\alpha$ and [\ion{O}{3}] lines across all scales, especially where imprints of the baryon fraction deviation exist, with space missions like SPHEREx. We show that although SPHEREx may only marginally probe $\mathcal{N}$ by accessing a modest number of large-scale modes in its 200\,deg$^2$ deep survey, future infrared all-sky surveys reaching a comparable depth with an improved spectral resolution ($R \gtrsim 400$) are likely to constrain $\mathcal{N}$ to a precision of 10--30\%, sufficient for distinguishing models with varying feedback assumptions, out to $z\sim4$ using BAO intensity mapping. Leveraging this effect, large, cosmic-variance-limited LIM surveys in the far future can 
scrutinize the physical connection between galaxy evolution and the large-scale cosmological environment, while performing stringent tests of the standard cosmological model.  
\end{abstract}
%% Keywords should appear after the \end{abstract} command. 
%% See the online documentation for the full list of available subject
%% keywords and the rules for their use.
\keywords{cosmology: theory -- large-scale structure of universe -- galaxies: evolution}

% ========== Section 1: Introduction ========== %

\section{Introduction} \label{sec:intro}

The coupling between radiation and baryons prior to cosmic recombination drives primordial acoustic waves that leave characteristic imprints on the matter power spectrum through the gravitational effect of baryons. These so-called baryon acoustic oscillations (BAOs) exist on a typical scale of about $100\,$cMpc, and have become an important standard ruler in cosmology \cite[][]{Eisenstein_2005}. Baryons are later on coupled to dark matter through gravity and can be perceived as a (biased) tracer of the matter distribution, with a roughly constant bias factor on large, linear scales. However, a scale-dependent bias is predicted to be induced by fluctuations of the relative baryon fraction measured by the local densities of baryons and dark matter (\citealt{BL_2011}, hereafter \citetalias{BL_2011}; \citealt{Angulo_2013}; \citealt{Schmidt_2016}; \citealt{Soumagnac_2016}). Detection of this effect has been attempted through BAO measurements using galaxies from the SDSS-III BOSS data, in the aim of testing the standard cosmological paradigm and connecting the light-to-mass ratio of galaxies to their large-scale cosmological environment \citep{Soumagnac_2016, Soumagnac_2019}. However, the results remain inconclusive due to the limited sample size and imaging quality of SDSS data. 

Given the close connection between the BAO-induced modulation of the baryon fraction and the mass-to-light ratio of galaxies, the scale-dependent modulation is a useful probe of how the star formation activity is related to the gas content of galaxies --- a fundamental relation of galaxy evolution often referred to as the global star formation law \citep{Kennicutt_1998, Daddi_2010, Krumholz_2012, Liu_2015, dlR_2019, Kennicutt_2021}. Astronomical determination of it relies on accurately measuring multi-wavelength proxies of the ongoing star formation rate (SFR; e.g., rest-frame UV continuum) and the gas mass (e.g., CO line luminosity) from selected galaxy samples, and therefore tends to be demanding and susceptible to various systematics, such as dust obscuration, gas excitation, and selection bias, especially at $z>2$ \citep{Casey_2014}. To date, the state-of-art analysis is still restricted to a relatively small sample of several hundred nearby galaxies \cite[see][]{dlR_2019, Kennicutt_2021}, while yet more comprehensive analyses of larger sample sizes and/or at higher redshifts are limited by requirements for high-quality, multi-wavelength data. Statistical constraints from BAO amplitudes thus represent a novel independent way to characterize the global star formation law, including its potential redshift evolution and multi-modality \citep{Santini_2014, Kennicutt_2021}, without the necessity of detecting individual galaxies. 

A concept originating from the field of observational cosmology, the line intensity mapping (LIM) technique has received increasing attention in recent years as a powerful means to study the astrophysics of galaxies and the intergalactic medium \citep{Chang_2019BAAS, Kovetz_2019BAAS}. In particular, the tight connection between the emission-line production and the astrophysics of interstellar gas of galaxies makes LIM a promising statistical probe of galaxy evolution. Historically, large-scale fluctuations of line intensity fields have been mainly considered for cosmological applications, such as probing alternative dark matter models, dark energy, gravitational lensing, neutrino properties, and the primordial non-Gaussianity \cite[e.g.,][]{Sitwell_2014, Karkare_2018, Bernal_2019, LB_2021, Chung_2022, Maniyar_2022, AMD_2022}. The majority of astrophysical explorations of LIM have been focusing on small, non-linear scales, where astrophysical processes of galaxy evolution are manifested through their effects on the one-halo or shot-noise components of the LIM power spectrum \cite[e.g.,][]{Wolz_2017, BA_2019, MRC_2020, SW_2021}. Therefore, it is interesting to extend the scope of astrophysical information from LIM to the linear regime by measuring baryon fraction fluctuations with BAO intensity mapping. The wide-bandwidth and coarse-grain averaged nature of LIM also makes it convenient to conduct coherent analysis of large statistical samples at multiple redshifts, thereby constraining any time evolution. 

LIM will be a main survey strategy of future space missions such as the Spectro-Photometer for the History of the Universe,
Epoch of Reionization, and Ices Explorer \cite[SPHEREx;][]{Dore_2014} and the Cosmic Dawn Intensity Mapper \cite[CDIM;][]{Cooray_2019}, which promise to offer unprecedented surface brightness sensitivity at near-infrared wavelengths that may allow to detect signals as faint as that expected from the first stars in the universe \cite[see e.g.,][]{Sun_2021, Parsons_2021}. High signal-to-noise measurements of large-scale LIM signals of optical/UV lines like H$\alpha$ 6563\,\AA, [\ion{O}{3}] 5007\,\AA\ and [\ion{O}{2}] 3727\,\AA\ are also made possible at $1 \lesssim z \lesssim 4$ \citep{Gong_2017}. It is therefore intriguing to understand how BAO imprints of the baryon fraction deviation may be utilized by these future experiments to study galaxy evolution at intermediate redshifts.

In this paper, we propose a novel method of constraining the global star formation law of galaxies from cosmological measurements of BAO intensity mapping. In Section~\ref{sec:models}, we present the modeling framework of the line intensity field in the presence of baryon fraction fluctuations and how LIM surveys can leverage the scale-dependent bias induced on BAO scales to extract the star formation law power index. In Section~\ref{sec:obs}, we use SPHEREx as an example to investigate the observational prospects for our method and forecast the constraining power on the parameters of interest based on the estimated detectability of H$\alpha$ and [\ion{O}{3}] LIM signals. We discuss some limitations and future improvements of our analysis, before concluding in Section~\ref{sec:summary}. 

Throughout the paper, all physical quantities related to star formation are normalized to have a \citet{Chabrier_2003} initial mass function, and we assume a flat, $\Lambda$CDM
cosmology consistent with the results from \citet{Planck_2016A&A}.

\section{Models} \label{sec:models}

The BAO-induced modulation of the relative clustering of baryons and dark matter is proposed by \citetalias{BL_2011} as a useful cosmological probe. In this work, we reformulate the original observational proposal, which involves measurements of the (original and luminosity-weighted) number density fields of galaxies, into a framework of multi-tracer LIM observations, which avoids the complications associated with flux-limited samples and alleviates parameter degeneracies. 

\subsection{Intensity Fields and Halo Baryon Fraction} \label{sec:sec:baryon_fraction}

\subsubsection{Perturbations of Halo Baryon Fraction}
Here we briefly review the physical concepts behind the halo baryon fraction perturbations on BAO scales, and interested readers are referred to \citetalias{BL_2011} for more details. 

The scale-dependent modulation of the baryon and dark matter density fields due to primordial acoustic waves can be described as
\begin{equation}
\delta_\gamma = \delta_\mathrm{b} - \delta_\mathrm{tot} = r\delta_\mathrm{tot},
\end{equation}
where $\delta_\mathrm{tot}$ is the total matter overdensity. We use CAMB \citep{Lewis_2000} to obtain $r(k,z)=\delta_\mathrm{b}/\delta_\mathrm{tot}-1$, the fractional deviation of the \textit{global} baryon fraction $\gamma_b$ whose cosmic mean value is $\bar{\gamma}_\mathrm{b} \equiv \Omega_\mathrm{b}/\Omega_\mathrm{m}$. Relatedly, the \textit{halo} baryon fraction is $f_\mathrm{b} = (\delta_\mathrm{b}/\delta_\mathrm{tot})\bar{\gamma}_\mathrm{b}=[1+r(k,z)]\bar{\gamma}_\mathrm{b}$. By separating perturbations into large-scale and small-scale effects of the density field and halo collapse, the lowest-order perturbation of the halo baryon fraction $f_\mathrm{b}$ is
\begin{equation}
\delta_f = \frac{A_r}{\delta_\mathrm{c}} \left[r(k)-r_\mathrm{LSS}\right] \delta_\mathrm{tot},
\end{equation}
where $\delta_\mathrm{c}=1.686$ is the critical density for spherical collapse in linear theory. The mostly constant, small-scale baryon fraction deviation, $r_\mathrm{LSS}$, can be well-approximated by $r(k=1\,h\,\mathrm{Mpc}^{-1})$, and $A_r \approx 3$ is a constant factor describing the enhancement of gas depletion into halos due to non-linear collapse, which can be characterized by simulations \citep{Naoz_2011,Angulo_2013}.

\subsubsection{Connection to Line Intensity Fields} \label{sec:model:connection_to_if}
For a given line tracer of the LSS with line luminosity $L$, we have \citepalias{BL_2011}
\begin{equation}
\delta_{L} = b_{L}(k) \delta_{\rm tot} = \left\{ b_{L,\mathrm{eff}} + b_{L;\Delta} [r(k)-r_\mathrm{LSS}] \right\} \delta_{\rm tot},
\label{eq:total_b}
\end{equation}
where $b_{L,\mathrm{eff}}$ is the luminosity-weighted effective bias of the tracer stems from the halo bias\footnote{Because we work solely with LIM signals in this work, bias contributions from the source number density and the luminosity weighting considered separately in \citetalias{BL_2011} are combined into a single $b_{L,\mathrm{eff}}$.}, which is taken to be scale-independent, and $b_{L;\Delta}=(A_r/\delta_\mathrm{c}) \mathcal{N} \beta$ denotes the additional, scale-dependent bias associated with perturbations of $f_\mathrm{b}$. Factors $\mathcal{N}$ (the star formation power law index) and $\beta$ (the $L$--SFR power law index) for the luminosity weighting of $f_\mathrm{b}$ are determined by galaxy astrophysics. Both observations and analytic models invoking feedback regulations suggest a universal, power-law relation between star formation activity and the gas content of galaxies, namely the global star formation law. Specifically, the star formation rate surface density is related to the gas surface density by $\dot{\Sigma}_* \propto (\Sigma_\mathrm{g})^{\mathcal{N}} \propto (f_\mathrm{b})^{\mathcal{N}}$, where the exact value of $\mathcal{N}$ is sensitive to astrophysical processes like stellar feedback \citep{Dekel_2019}. For example, the well-known Kennicutt-Schmidt law suggests $\mathcal{N} \approx 1.4$, whereas \citet{FQH_2013} propose a simple model where the galaxy disc is supported entirely by stellar feedback and find $\mathcal{N} \approx 2$. If the $L$--SFR relation also follows a simple power law with index $\beta$ (as is usually the case, see Section~\ref{sec:sec:sec:lsfr}), then
\begin{equation}
L \propto \dot{M}_{*}^{\beta} \propto (f_\mathrm{b})^{\mathcal{N}\beta},
\label{eq:bLfb}
\end{equation}
which implies $b_{L;\Delta} \propto \mathcal{N}\beta$, as discussed above. 

\begin{figure}[h]
 \centering
 \includegraphics[width=0.47\textwidth]{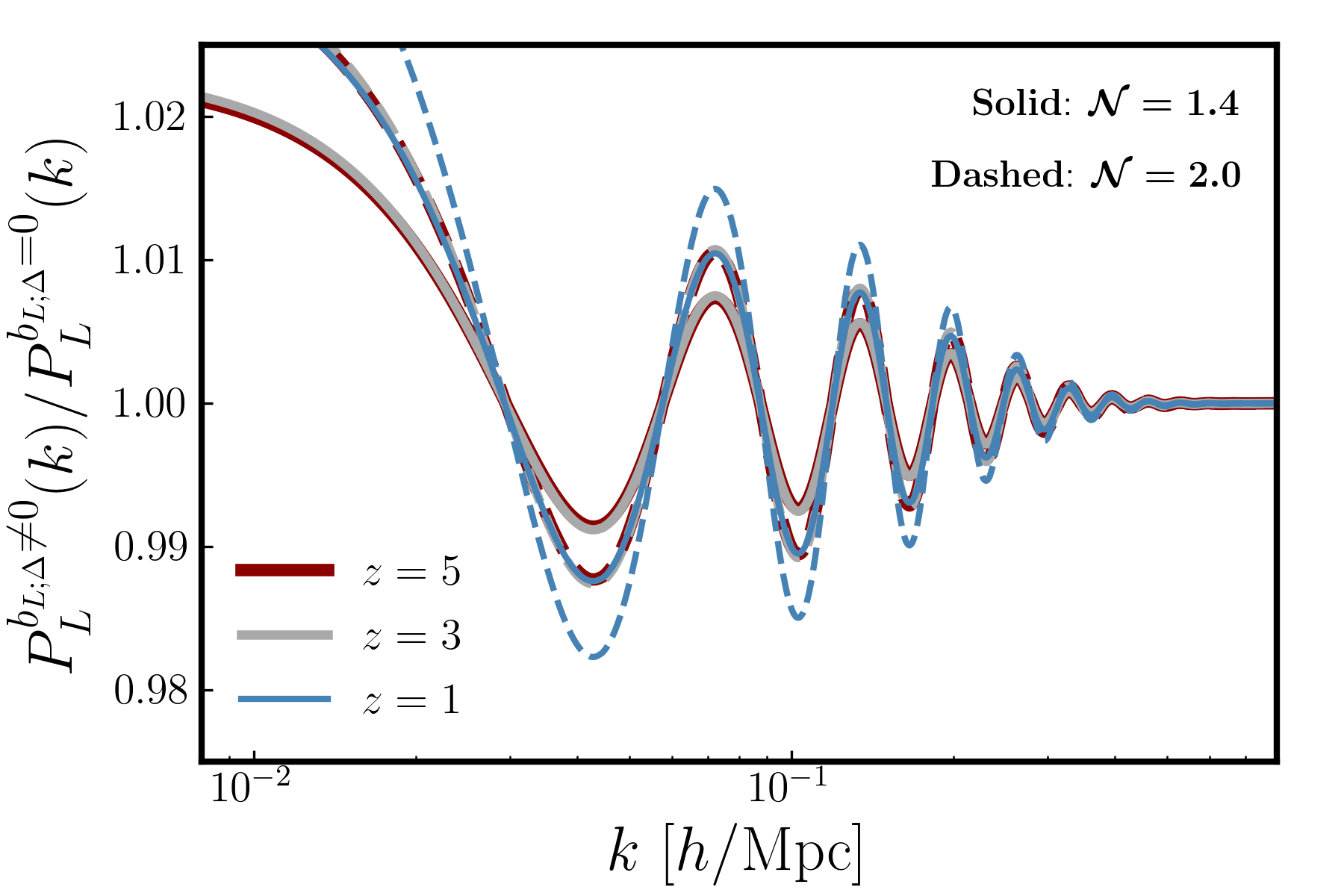}
 \caption{The ratio of the line intensity power spectrum with and without accounting for the scale-dependent bias of a typical LSS tracer with $\beta=1$ due to baryon fraction fluctuations. The solid and thin curves represent two familiar forms of the star formation law, corresponding to the Kennicutt--Schmidt law \cite[$\mathcal{N}\approx1.4$,][]{Kennicutt_1998} and a simple feedback-supported disc model \cite[$\mathcal{N}\approx2$,][]{FQH_2013}, respectively.}
 \label{fig:psbias}
\end{figure}

\citetalias{BL_2011} make an initial observational proposal to use the scale dependence from $r(k)$, which modulates the BAO peak amplitudes but barely changes the peak positions, to separate different bias contributions. Specifically, they propose to compare the number density and luminosity density power spectra of the same galaxy sample to cancel out sources of foreground contamination in common. By analogy, we note that LIM data enable a major simplification of observables thanks to their sensitivity to the aggregate line emission \citep{VL_2010}. The (square-rooted) power spectrum ratio of two LIM signals is
\begin{equation}
\mathcal{R} \equiv \left(\frac{P_1}{P_2}\right)^{1/2} = \mathcal{B}_1 \left\{ 1 + \mathcal{B}_2 [r(k)-r_\mathrm{LSS}] \right\}, 
\label{eq:Rdef}
\end{equation}
where we assume the large-scale limit and that the two line tracers share the same $b_{L;\Delta}$ (but see Section~\ref{sec:models:ps} for a full treatment), such that the coefficients can be expressed as $\mathcal{B}_1 = b_{L_1,\mathrm{eff}}/b_{L_2,\mathrm{eff}}$ and $\mathcal{B}_2=(b_{L;\Delta}/b_{L_2,\mathrm{eff}})(\mathcal{B}^{-1}_1-1)$, respectively. Similar to what \citetalias{BL_2011} find, we see that the scale dependence of $r(k)$ allows to separately constrain $\mathcal{B}_1$ and $\mathcal{B}_2$ with $\mathcal{R}$, and thereby infer $b_{L;\Delta}$. Nonetheless, several important issues are apparent. First, with $\mathcal{R}$ alone, it is clearly infeasible to decouple individual bias factors $b_{L_1,\mathrm{eff}}$, $b_{L_2,\mathrm{eff}}$, and $b_{L;\Delta}$ from their ratios. Second, to deduce the global star formation law represented by $\mathcal{N}$ from $b_{L;\Delta}$, the parameter $\beta$ must be known a priori, though it can actually vary significantly for a single line tracer under different astrophysical conditions, or for different tracers as is relevant here. 

Therefore, to ultimately constrain $\mathcal{N}$, some means in addition to the measurement of $\mathcal{R}$ is needed to lift the degeneracies among different bias factors and account for astrophysical uncertainties in the $L$--SFR relation. In what follows, we investigate and demonstrate a natural extension of Equation~(\ref{eq:Rdef}) in the context of multi-tracer LIM. The cross-correlation between the two line tracers and the full shape of the line intensity power spectrum including the small-scale, shot-noise term are incorporated into the analysis, in order to maximally separate the different bias factors and astrophysical parameters. 

\subsection{LIM Observables and Emission Line Models}

\subsubsection{Power Spectrum and Galaxy--Halo Connection} \label{sec:models:ps}

For a given line $L$, the power spectrum of line intensity fluctuations is
\begin{equation}
P_{L}(k) = \langle I_L \rangle^2 b^2_{L}(k) P_{\delta \delta}(k) + P_{L,\mathrm{shot}},
\end{equation}
where $b_{L}(k)$ is the net bias factor of the tracer defined in Equation~(\ref{eq:total_b}) and $P_{\delta\delta}(k)$ is the matter power spectrum of $\delta_\mathrm{tot}$. Scale-independent factors $\langle I_L \rangle$ and $P_{L,\mathrm{shot}}$ represent the mean line intensity and the shot-noise power arising from the Poissonian distribution of discrete line emitters, respectively. We compute them from the $L$--SFR relation as
\begin{equation}
\langle I_L \rangle = \int dM \frac{dn}{dM} \frac{L[\dot{M}_*(M,z)] y(z) D^2_A}{4\pi D^2_L},
\end{equation}
and
\begin{equation}
P_{L,\mathrm{shot}} = \int dM \frac{dn}{dM} \left\{\frac{L[\dot{M}_*(M,z)] y(z) D^2_A}{4\pi D^2_L} \right\}^2, 
\end{equation}
where $\langle I_L \rangle$ is related to the line luminosity density by $y(z)=\lambda_\mathrm{obs}(1+z)/H(z)$, the comoving angular diameter distance $D_A$, and the luminosity distance $D_L$. We adopt the halo mass function $dn/dM$ from \citet{Tinker_2008} and evaluate the integrals over $10^9\,M_{\odot} < M < 10^{15}\,M_{\odot}$. Figure~\ref{fig:psbias} shows how the line intensity power spectra compare with and without including the scale-dependent bias induced by baryon fraction fluctuations. 

For simplicity, we assume a one-to-one correspondence between halos and galaxies, and ignore details of the halo occupation distribution (HOD) and the stochasticity in the line luminosity that can have non-trivial effects on small scales \cite[see e.g.,][]{Sun_2019, SW_2021}. For each galaxy, we obtain its SFR, $\dot{M}_*(M,z)$, from the Data Release 1 of the \texttt{UniverseMachine} code \citep{Behroozi_2019}, which semi-empirically models the correlated halo assembly and galaxy growth. 

\begin{figure}[h]
 \centering
 \includegraphics[width=0.47\textwidth]{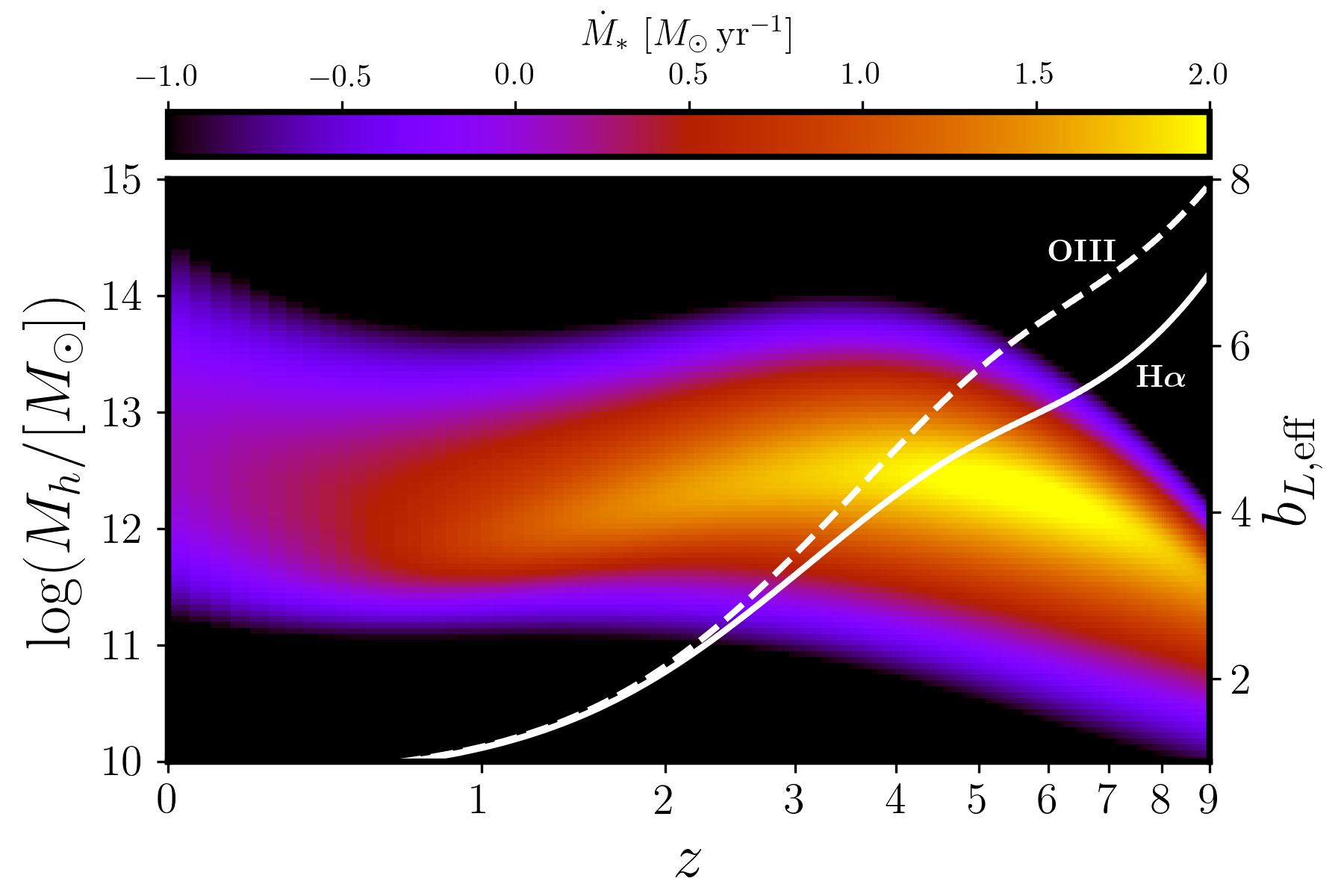}
 \caption{The SFR as a function of halo mass and redshift obtained from the \texttt{UniverseMachine} code \citep{Behroozi_2019}. Also plotted are the implied luminosity-weighted effective bias factors of the two tracers, $b_{\mathrm{H\alpha, eff}}(z)$ and $b_{\mathrm{OIII, eff}}(z)$, as labeled by the right axis. }
 \label{fig:sfr_and_bias}
\end{figure}

\subsubsection{L--SFR Relation} \label{sec:sec:sec:lsfr}

As discussed in Section~\ref{sec:sec:baryon_fraction}, the line intensity serves as a biased tracer of the local matter density, which is subject to not only the halo occupation and environmental dependence of galaxy evolution but also the relation between the production of line photons and the baryon fraction of galaxies. While $b_{L, \mathrm{eff}}$ accounts for the fluctuations sourced by the dependence of source number density and luminosity on the local matter density, we also need to specify the $L$--SFR relation to model $b_{L;\Delta}$. Motivated by the well-established correlation observed between the SFR and the luminosity of lines as star formation tracers, we take 
\begin{equation}
\log (L/[\mathrm{erg\,s^{-1}}]) = \alpha + \beta \log (\dot{M}_{*}/[M_{\odot}\,\mathrm{yr^{-1}}]),
\label{eq:lsfr}
\end{equation}
where for each line we vary both $\alpha$ (affecting the power spectrum amplitude) and $\beta$ (affecting both the power spectrum amplitude and shape). In Figure~\ref{fig:sfr_and_bias}, we plot over the $\dot{M}_*(M,z)$ space the scale-independent, effective bias of two promising target lines for LIM, H$\alpha$ and [\ion{O}{3}], that will be studied by SPHEREx. Note that while we treat $b_{L,\mathrm{eff}}$ as a free parameter in our Fisher matrix analysis (see Section~\ref{sec:obs}), we use fiducial values of $\alpha$ and $\beta$ to obtain the fiducial $b_{L,\mathrm{eff}}$ of our tracers from the scale-independent halo bias. As shown in Figure~\ref{fig:sfr_and_bias}, $b_{L,\mathrm{eff}}$ of either line evolves by about a factor of 5 and the different redshift trends are associated with the different fiducial $\beta$ values taken (see Table~\ref{tb:model_params}).  

\begin{table}[h!]
\caption{Fiducial model parameters and priors}
\label{tb:model_params}
\centering
\begin{tabular}{cccc}
\toprule
\toprule
Parameter & Input Value ($z=1$) & Prior & Reference \\
\hline
$b_{\mathrm{H\alpha, eff}}$ & 1.2 & $100\%$ & Eq.~(\ref{eq:total_b}) \\
$b_{\mathrm{OIII, eff}}$ & 1.2 & $100\%$ & Eq.~(\ref{eq:total_b}) \\
$\mathcal{N}$ & 1.4 & $100\%$ & Eq.~(\ref{eq:bLfb}) \\
$\alpha_{\mathrm{H\alpha}}$ & 41.1 & $100\%$ & Eq.~(\ref{eq:lsfr}) \\
$\beta_{\mathrm{H\alpha}}$ & 1.0 & $10\%$ & Eq.~(\ref{eq:bLfb}, \ref{eq:lsfr}) \\
$\alpha_{\mathrm{OIII}}$ & 41.0 & $100\%$ & Eq.~(\ref{eq:lsfr}) \\
$\beta_{\mathrm{OIII}}$ & 1.2 & $50\%$ & Eq.~(\ref{eq:bLfb}, \ref{eq:lsfr}) \\
%$\langle I_{\mathrm{H\alpha}} \rangle\ (\mathrm{Jy/sr})$ & 19.4, 5.8, 1.4 & $\cdots$ & Eq.~(\ref{eq:bLfb}) \\
%$\langle I_{\mathrm{OIII}} \rangle\ (\mathrm{Jy/sr})$ & 15.9, 5.7, 1.1 & $\cdots$ & Eq.~(\ref{eq:bLfb}) \\
\bottomrule
\end{tabular}
\end{table}

\section{Observational Prospects} \label{sec:obs}

\subsection{Basic Setups} \label{sec:obs:basic}

To demonstrate the capability of BAO intensity mapping for probing the global star formation law through the baryon fraction deviation, in this section we envisage a case study to jointly measure H$\alpha$ and [\ion{O}{3}] LIM signals at $z=1$--5 with $\Delta z = 0.5$ using a SPHEREx-like experiment. 
We empirically model the LIM signals using the best-fit results to the observed $L$--SFR relations, assuming $\alpha_{\mathrm{H}\alpha}=41.1$ and $\beta_{\mathrm{H}\alpha}=1.0$ for H$\alpha$ \citep{Ly_2007}, and $\alpha_{\mathrm{OIII}}=41.0$ and $\beta_{\mathrm{OIII}}=1.2$ for [\ion{O}{3}] \citep{VillaVelez_2021}. Table~\ref{tb:model_params} summarizes the fiducial input values and priors (1$\sigma$, quoted in percentage of the fiducial value) of model parameters, on which the constraints from mock observations are estimated through the Fisher matrix analysis. 

As discussed in Section~\ref{sec:model:connection_to_if}, to explore the parameter constraints that SPHEREx-like experiments can provide, we investigate the observational prospects for measuring together the ratio of H$\alpha$ and [\ion{O}{3}] auto-power spectra, $\mathcal{R} = \sqrt{P_{\mathrm{OIII}}/P_{\mathrm{H}\alpha}}$, and their cross-power spectrum, $\mathcal{P} = P_{\mathrm{OIII}\times\mathrm{H\alpha}}$. There are two main reasons that we choose $\mathcal{R}$ instead of using the auto-power spectra of the respective lines. It preserves the format of the metric proposed in \citetalias{BL_2011}, which can be easily separated into scale-independent and scale-dependent terms. More importantly, given that auto-power spectra are often contaminated by some common sources of foreground such as the atmospheric emission and the extragalactic background light, taking the ratio makes it more justified to use the (propagated) signal-to-noise ratio (S/N) estimated from simple mode counting. 

We then adopt the survey specifications of SPHEREx to estimate the detectability of $\mathcal{R}$ and $\mathcal{P}$, following procedures outlined in e.g., \citet{Gong_2017}. While the all-sky survey of SPHEREx is more advantageous for measuring the BAO amplitudes on large scales, it is too shallow compared to the 200\,deg$^2$ SPHEREx deep survey, which is approximately 7 times deeper in terms of the surface brightness sensitivity and thus more suitable for LIM applications. Thus, we assume a survey area of 200\,deg$^2$ and a spectral resolving power of $R=40$, consistent with the way H$\alpha$ and [\ion{O}{3}] LIM will be conducted by SPHEREx in its four shortest-wavelength bands. There are, however, two noteworthy differences from \citet{Gong_2017}. First, for the surface brightness sensitivity that determines the instrument noise power, $P_\mathrm{n}$, we assume the current best estimate (CBE) performance of SPHEREx\footnote{See the public product released at \url{https://github.com/SPHEREx/Public-products/blob/master/Surface_Brightness_v28_base_cbe.txt}}, which leads to an approximately 100 times lower $P_\mathrm{n}$. Second, given the relatively low spectral resolution of SPHEREx, we include an extra smoothing factor, $G(k)$, in the sensitivity calculation that accounts for the attenuation of the signal power spectrum on small scales due to finite spatial and spectral resolutions. Although our baseline model predicts H$\alpha$ and [\ion{O}{3}] signal levels similar to those in \citet{Gong_2017} and that the total S/N estimates of the power spectra differ only by a factor of 2, the two aforementioned factors imply S/N distributions as a function of $k$ much different from \citet{Gong_2017}. 

\begin{figure}
 \centering
 \includegraphics[width=0.47\textwidth]{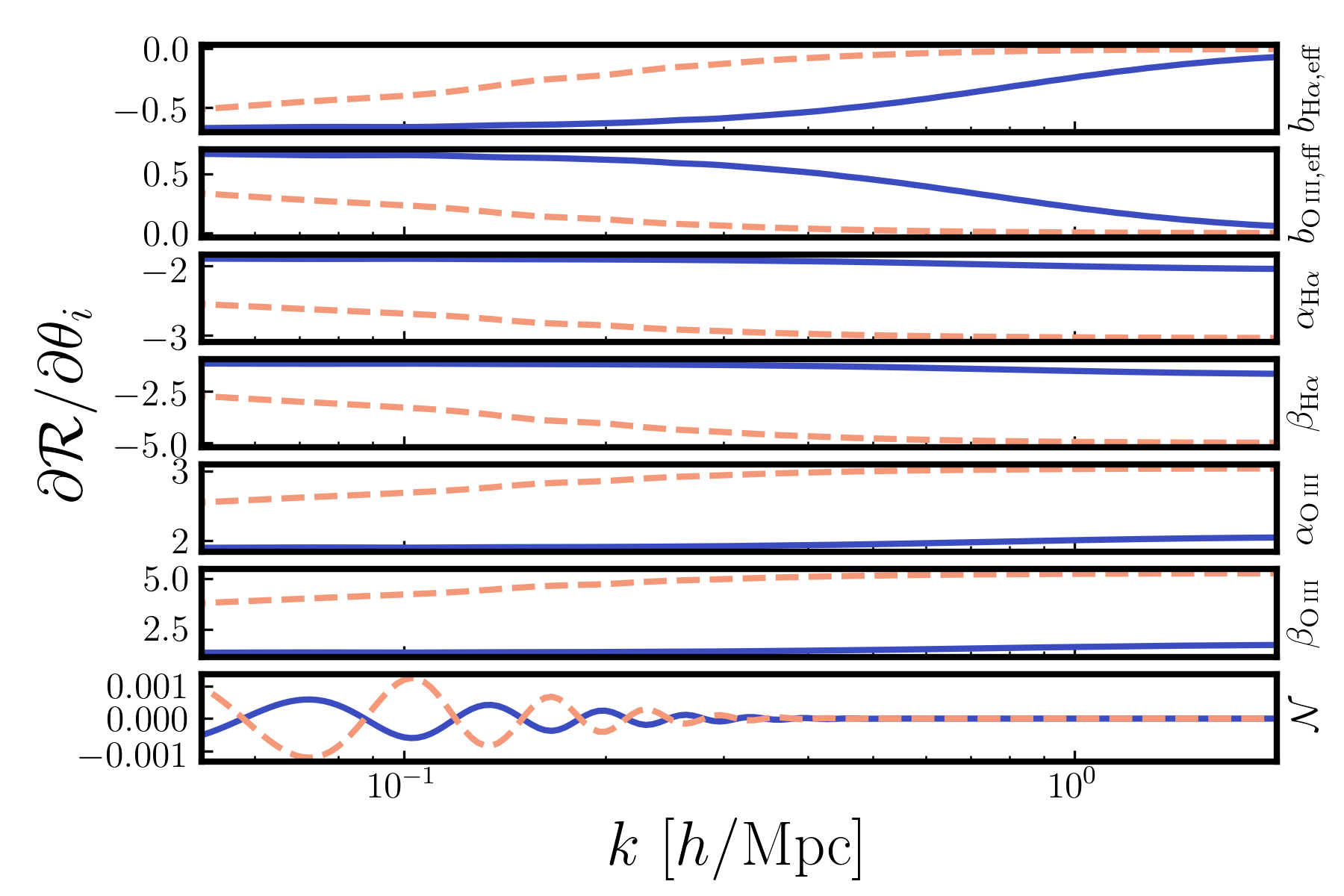}
 \includegraphics[width=0.47\textwidth]{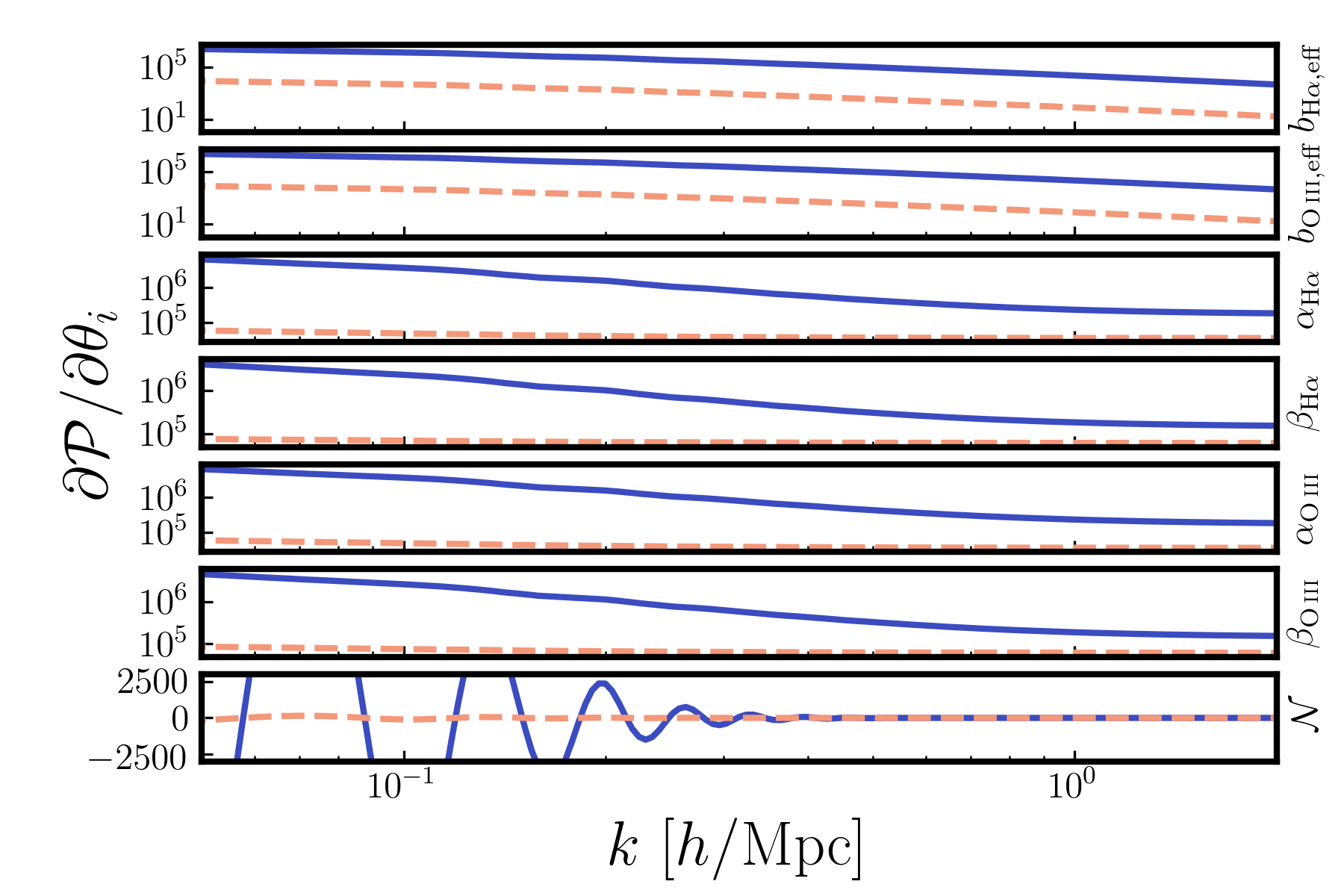}
 \caption{Parameter derivatives of the two observables, $\mathcal{R}$ and $\mathcal{P}$, entering the Fisher matrix analysis. From top to bottom, the 7 panels show the derivatives with respect to $b_\mathrm{H\alpha, eff}$, $b_\mathrm{O\,\textsc{iii}, eff}$, $\alpha_\mathrm{H\alpha}$, $\beta_\mathrm{H\alpha}$, $\alpha_\mathrm{O\,\textsc{iii}}$, $\beta_\mathrm{O\,\textsc{iii}}$, and $\mathcal{N}$, respectively, as a function of the wavenumber at $z=1$ (blue solid) and $z=4$ (red dashed).}
 \label{fig:fisherderivs}
\end{figure}

\subsection{Fisher Matrix Analysis}

\begin{figure*}
 \centering
 \includegraphics[width=0.85\textwidth]{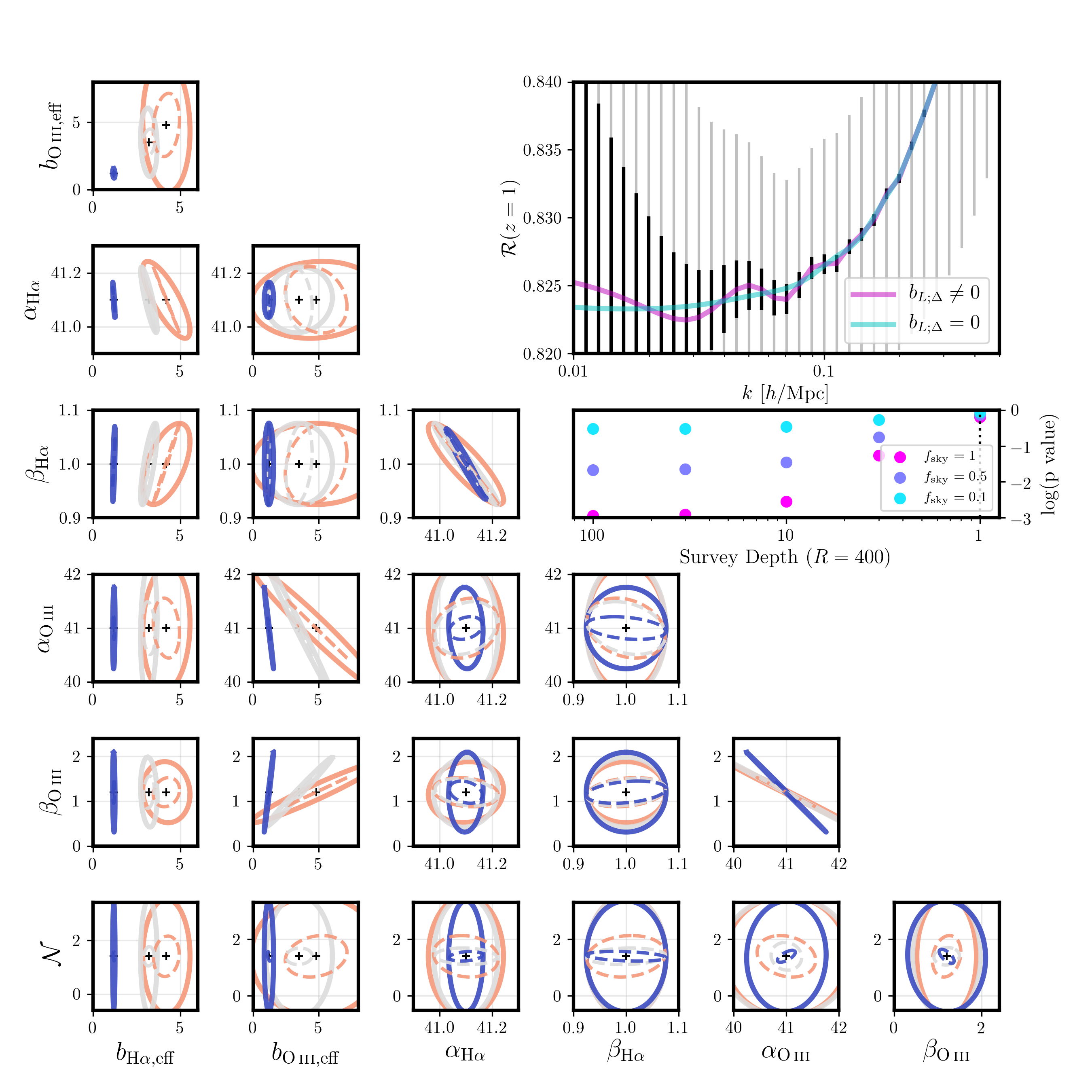}
 \caption{Constraints on the model parameters and their degeneracies from the Fisher matrix analysis performed at $z=1$ (blue), 3 (gray), and 4 (red). The black cross indicates the true input value, with multiple values shown for $b_{L, \mathrm{eff}}$ which increases with redshift. The solid and dashed contours represent constraints from the 200\,deg$^2$ SPHEREx deep fields and a hypothetical all-sky survey reaching the same depth but with $R=400$, whose constraining power on the power spectrum ratio $\mathcal{R}$ at $z=1\pm0.5$ are shown by the error bars in gray and black in the upper inset, respectively. Note that uncertainties of the SPHEREx deep survey are reduced by 3 times to aid comparison. Models with and without the BAO-induced scale-dependent bias are shown by the magenta and cyan curves, respectively, for comparison. The lower inset shows the distinguishing power (in p value corresponding to the chi-square difference) between models with and without the scale-independent bias, as a function of the sky coverage $f_{\rm sky}$ and the survey depth compared with the nominal depth of the SPHEREx all-sky survey (vertical dotted line), measured by the surface brightness sensitivity.}
 \label{fig:fishercorner}
\end{figure*}

With estimates of the target observables and SPHEREx sensitivities in hand, we calculate the covariance matrix of the model parameters using the Fisher matrix
\begin{equation}
    F_{ij} = \sum_k \frac{1}{\mathrm{var}(f)} \frac{\partial f}{\partial \theta_i} \frac{\partial f}{\partial \theta_j},
\end{equation}
where the summation is over all the $k$ bins for the data vector $f(\theta, k)=(\mathcal{R}(\theta, k), \mathcal{P}(\theta, k))$, and the covariance between the two observables is neglected since auto- and cross-power measurements are subject to generally uncorrelated systematic uncertainties. 

Figure~\ref{fig:fisherderivs} shows the parameter derivatives of the two observables entering the Fisher matrix analysis, $\mathcal{R}$ and $\mathcal{P}$. The way they are modulated by the model parameters can be perceived from the shape of the curves, which characterizes the scale dependence of the modulation. For instance, from both $\partial{\mathcal{R}}/\partial{b_{L, \mathrm{eff}}}$ and $\partial{\mathcal{P}}/\partial{b_{L, \mathrm{eff}}}$, it is clear that $b_{L, \mathrm{eff}}$ has a diminishing effect on the observables towards smaller scales, which become increasingly dominated by the shot noise that only depends on $\alpha$ and $\beta$. For $\mathcal{N}$, because it only imposes small perturbations on the power spectrum amplitude on BAO scales, the derivatives with respect to it have qualitatively different shapes compared with others. Such distinctions in the scale dependence are key to the capability for separately constraining all free parameters with the two observables. Indeed, as suggested by the close similarity between some curves of derivatives, e.g., with respect to $\alpha$ and $\beta$, strong (anti-)correlation and thus degeneracy exist between these parameters. 

We show in Figure~\ref{fig:fishercorner} the projected constraints on all the free parameters together with their degeneracy patterns from our Fisher matrix analysis. Overall, jointly measuring $\mathcal{R}$ and $\mathcal{P}$ of H$\alpha$ and [\ion{O}{3}] at high significance makes it possible to robustly constrain the model parameters, including $\mathcal{N}$. Unfortunately, although  measurements of the observables are already cosmic-variance limited for the SPHEREx deep survey, it does not have large enough sky coverage and spectral resolution to measure a sufficient number of large-scale modes. With the assumed priors, $\mathcal{N}$ can only be measured to a 80\% precision with SPHEREx as shown by solid contours, with insufficient evidence for a scale-dependent bias induced by baryon fraction fluctuations, which we quantify by the p value corresponding to the chi-square difference between best-fit models with and without the scale-dependent bias $b_{L;\Delta}$ (a p value $\approx$ 0.9 is obtained in this case). While jointly fitting all redshift bins neglecting any redshift evolution improves the constraints, it is still hard to achieve meaningful constraints with the limited size and spectral resolution of the SPHEREx deep survey.

Thus, we also consider a more idealized all-sky survey with $R=400$ that provides a 10-fold increase in the number of accessible modes required to beat down the sample variance and the same survey depth as the SPHEREx deep survey. With similar instrument specifications but 10 times higher spectral resolution, a much lower system temperature (5\,K vs. 80\,K for SPHEREx) must be reached via active cooling to achieve a reasonable mission duration of about 2 years. As shown by the dashed contours, such a deep, all-sky survey allows the model parameters to be measured a lot more precisely and the constraints are much less prior-dominated. In this idealized case, a strong evidence for BAO-induced scale-dependent bias is observed (p value $\approx$ 0.005), and $\mathcal{N}$ can be determined to a precision of 10--30\% up to $z\sim4$, which allows to investigate the physical origin of the global star formation law and reveal any dependence on redshift or the galaxy population. For reference, in the upper inset of Figure~\ref{fig:fishercorner}, we show a comparison of the constraints on $\mathcal{R}$ from the two surveys considered, together with best-fit models with and without introducing the scale-independent bias. In the lower inset, we show how the p value of chi-square difference between the best-fit models changes with $f_\mathrm{sky}$ and the survey depth with respective to the SPHEREx all-sky survey. Consistent with what the dashed contours imply, an all-sky survey reaching the SPHEREx deep survey depth with $R=400$ is required for obtaining a strong evidence for the scale-dependent bias. 

Two other features are noteworthy from the resulting constraint ellipses. First, the degeneracy patterns displayed are generally well-expected from how the model parameters affect the two observables. Clear (anti-)correlations are evident between $\alpha_\mathrm{H\alpha}$ and $\beta_\mathrm{H\alpha}$, $\alpha_{\mathrm{O}\,\textsc{iii}}$ and $\beta_{\mathrm{O}\,\textsc{iii}}$, etc. Second, we do see a change of degeneracy direction between $\mathcal{N}$ and other parameters from $z=1$ to $z=3$ and 4. This is associated with a change in the dependence of the [\ion{O}{3}]--H$\alpha$ power ratio $\mathcal{R}$ on $\mathcal{N}$, which can be easily seen from the derivative curve of $\partial{\mathcal{R}}/\partial{\mathcal{N}}$ shown in the top panel of Figure~\ref{fig:fisherderivs}, due to the presence of shot-noise contribution $P_{L,\mathrm{shot}}$ in $\mathcal{R}$, which can alter the way a nonzero $\mathcal{N}$ impacts $\mathcal{R}$ at sufficiently high redshifts like $z=4$. 

\section{Discussion and Conclusions} \label{sec:summary}

So far, we have assessed how imprints of the baryon fraction deviation on BAO scales can be utilized by future LIM surveys to constrain the fundamental relationship between star formation and the gas content of galaxies. A number of caveats need to be noted though regarding our analysis. First, in practice, LIM data sets ultimately need to be analyzed allowing both astrophysics and cosmology to vary. This is particularly true for large-scale signals such as the BAOs considered in this work, and will inevitably make the extraction and interpretation of astrophysical information like $\mathcal{N}$ more challenging. Meanwhile, although we choose to leave them out of this work for succinctness, observational effects that complicate the target LIM signals, such as redshift-space distortions (RSDs) and line interlopers, are important factors to be accounted for in the actual data analysis. Fortunately, with techniques such as measuring the full multipole moments of redshift-space power spectrum, it is possible to reliably constrain both the astrophysics and cosmology, with effects of RSDs and interloping lines properly included \cite[see e.g.,][and references therein]{Gong_2020}. Finally, even on linear scales, astrophysical processes other than what the star formation law encodes may also introduce scale-dependent bias that can further complicate the interpretation of observations, for either astrophysics or cosmology. Some examples of such large-scale modulations include feedback \citep{CE_2007}, radiative transfer effects \citep{Pontzen_2014}, and the impact of galaxy formation physics on halo occupation statistics \citep{Angulo_2014}. 

In summary, the BAO-induced scale-dependent bias associated with baryon fraction fluctuations provides a useful way to probe astrophysics such as the global star formation law of galaxies on cosmological scales. Our analysis shows that LIM promises to measure this effect and directly constrain the global star formation law power index $\mathcal{N}$, using large-number statistics in a huge cosmic volume rather than zoom-in analyses of individual galaxies. However, such measurements are challenging to make, typically requiring an immense number of modes to achieve a high sensitivity to the BAO amplitudes, which is beyond the capability current-generation surveys like SPHEREx. Future all-sky LIM surveys reaching similar depth but with $\sim10$ times better spectral resolving power than SPHEREx will be capable of measuring $\mathcal{N}$ at high significance with BAO intensity mapping. Beyond performing stringent tests on the standard cosmological model, results from such surveys will examine in detail the galaxy evolution theory against the backdrop of large-scale structure formation. 

\acknowledgments

We thank Jordan Mirocha, Tzu-Ching Chang, Llu\'{i}s Mas-Ribas, and Jamie Bock for helpful conversations and comments, as well as the anonymous referee for comments that improved the manuscript. We acknowledge the support from the JPL R\&TD strategic initiative grant on line intensity mapping. 

%% To help institutions obtain information on the effectiveness of their 
%% telescopes the AAS Journals has created a group of keywords for telescope 
%% facilities.
%
%% Following the acknowledgments section, use the following syntax and the
%% \facility{} or \facilities{} macros to list the keywords of facilities used 
%% in the research for the paper.  Each keyword is check against the master 
%% list during copy editing.  Individual instruments can be provided in 
%% parentheses, after the keyword, but they are not verified.

%% Similar to \facility{}, there is the optional \software command to allow 
%% authors a place to specify which programs were used during the creation of 
%% the manuscript. Authors should list each code and include either a
%% citation or url to the code inside ()s when available.

%\vspace{5mm}
%\software{astropy \citep{2013A&A...558A..33A}, % 
%          Cloudy \citep{2013RMxAA..49..137F}, 
%          SExtractor %\citep{1996A&AS..117..393B}
%          }

%% Appendix material should be preceded with a single \appendix command.
%% There should be a \section command for each appendix. Mark appendix
%% subsections with the same markup you use in the main body of the paper.

%% Each Appendix (indicated with \section) will be lettered A, B, C, etc.
%% The equation counter will reset when it encounters the \appendix
%% command and will number appendix equations (A1), (A2), etc. The
%% Figure and Table counter will not reset.

\bibliography{baoim}{}
\bibliographystyle{aasjournal}

%% This command is needed to show the entire author+affiliation list when
%% the collaboration and author truncation commands are used.  It has to
%% go at the end of the manuscript.
%\allauthors

%% Include this line if you are using the \added, \replaced, \deleted
%% commands to see a summary list of all changes at the end of the article.
%\listofchanges

\end{document}